\documentclass[a4paper,fleqn,11pt]{article}
\pdfoutput=1 
\usepackage{amsmath}
\usepackage{feynmp-auto}
\usepackage{jheppub} 
\usepackage{mathrsfs}                     
\usepackage[T1]{fontenc} 
\usepackage{slashed}

\title{\boldmath Symmetries and unitary interactions of mass dimension one fermionic dark matter}

\author{Cheng-Yang Lee}

\affiliation{Institute of Mathematics, Statistics and Scientific Computation,\\
Unicamp, 13083-859 Campinas, S\~{a}o Paulo, Brazil}

\emailAdd{cylee@ime.unicamp.br}

\abstract{The fermionic fields constructed from Elko have several unexpected properties. They satisfy the Klein-Gordon but not the Dirac equation and are of mass dimension one instead of three-half. Starting with the Klein-Gordon Lagrangian, we initiate a careful study of the symmetries and interactions of these fermions and their higher-spin generalisations. We find, although the fermions are of mass dimension one, the four-point fermionic self-interaction violates unitarity at high-energy. Therefore, it cannot be a fundamental interaction of the theory. It follows that for the spin-half fermions, the demand of renormalisability and unitarity forbids four-point interactions and only allows for the Yukawa interaction. For fermions with spin $j>\frac{1}{2}$, they have no renormalisable or unitary interactions. Since the theory is described by a Klein-Gordon Lagrangian, the interaction generated by the local $U(1)$ gauge symmetry which contains a four-point interaction, is excluded by the demand of renormalisability. In the context of the Standard Model, these properties make the spin-half fermions natural dark matter candidates.

}

\newcommand{\dual}[1]{\overset{{}^{{}^{\boldsymbol{\neg}}}}{\smash[t]{#1}}} 
\begin{document} 
\unitlength=1mm
\maketitle
\flushbottom

\def\p{\mbox{\boldmath$\displaystyle\mathbf{p}$}}
\def\r{\mbox{\boldmath$\displaystyle\mathbf{r}$}}
\def\e{\mbox{\boldmath$\displaystyle\mathbf{\epsilon}$}}
\def\th{\mbox{\boldmath$\displaystyle\mathbf{\theta}$}}
\def\k{\mbox{\boldmath$\displaystyle\mathbf{k}$}}
\def\g{\mbox{\boldmath$\displaystyle\mathbf{g}$}}
\def\q{\mbox{\boldmath$\displaystyle\mathbf{q}$}}
\def\bv{\mbox{\boldmath$\displaystyle\mathbf{\varphi}$}}
\def\hp{\mbox{\boldmath$\displaystyle\mathbf{\widehat{\p}}$}}
\def\0{\mbox{\boldmath$\displaystyle\mathbf{0}$}}
\def\O{\mbox{\boldmath$\displaystyle\mathbf{O}$}}
\def\s{\mbox{\boldmath$\displaystyle\mathbf{\sigma}$}}
\def\J{\mbox{\boldmath$\displaystyle\mathbf{J}$}}
\def\K{\mbox{\boldmath$\displaystyle\mathbf{K}$}}
\def\sJ{\mbox{\boldmath$\displaystyle\mathscr{J}$}}
\def\sK{\mbox{\boldmath$\displaystyle\mathscr{K}$}}
\def\mJ{\mbox{\boldmath$\displaystyle\mathcal{J}$}}
\def\iJ{\mbox{\boldmath$\displaystyle\mathit{J}$}}
\def\iK{\mbox{\boldmath$\displaystyle\mathit{K}$}}
\def\mK{\mbox{\boldmath$\displaystyle\mathcal{K}$}}
\def\x{\mbox{\boldmath$\displaystyle\mathbf{x}$}}
\def\y{\mbox{\boldmath$\displaystyle\mathbf{y}$}}
\def\bk{\mbox{\boldmath$\displaystyle\mathbf{\mathcal{K}}$}}
\def\bg{\mbox{\boldmath$\displaystyle\mathbf{\gamma}$}}

\section{Introduction}
The Standard Model of particle physics (SM) has been successful in describing the properties of elementary particles. But despite its success, there remains many outstanding problems in particle physics which led to the consensus that the theory is incomplete. Among them, one of the most important problem is the nature of non-baryonic dark matter~\cite{Hinshaw:2012aka,Bennett:2012zja,Ade:2015xua}. 



The theory of Elko and mass dimension one fermion presents an intriguing new paradigm for physics beyond the SM as well as for quantum field theory~\cite{Ahluwalia:2004ab,Ahluwalia:2004sz}. In particle physics, the fermion is a dark matter candidate while in quantum field theory, one has a new fermion with unexpected properties -- the field satisfies the Klein-Gordon but not the Dirac equation and is of mass dimension one instead of three-half. Since its original conception, the theory has established itself beyond particle physics and quantum field theory~\cite{Ahluwalia:2008xi,Ahluwalia:2009rh,Lee:2012td,Dias:2010aa,Alves:2014kta,Alves:2014qua,Agarwal:2014oaa,
Fabbri:2010va,Fabbri:2009aj,Fabbri:2009ka} into other disciplines. They have found applications in cosmology~\cite{Boehmer:2010ma,Boehmer:2010tv,Boehmer:2009aw,Boehmer:2008ah,Boehmer:2008rz,
Boehmer:2007ut,Boehmer:2007dh,Boehmer:2006qq,Chee:2010ju,Wei:2010ad,Shankaranarayanan:2010st,
Shankaranarayanan:2009sz,Gredat:2008qf,S.:2014dja,Pereira:2014wta,Basak:2014qea}, gravity~\cite{Jardim:2014xla,daRocha:2014dla,Fabbri:2014foa} and mathematical physics~\cite{HoffdaSilva:2009is,daRocha:2008we,daRocha:2007pz,daRocha:2005ti,daRocha:2011yr,
daRocha:2013qhu,Cavalcanti:2014wia,Cavalcanti:2014uta,Bonora:2014dfa,Ablamowicz:2014rpa}.

One of the most important objective for the theory is to establish a rigorous platform to study the phenomenologies of these fermions. On this front, the particle signatures at the Large Hadron Collider and in astrophysics have been studied~\cite{Dias:2010aa,Alves:2014kta,Alves:2014qua,Agarwal:2014oaa}. However, to the best of our knowledge, these studies did not consider all the possible interactions. In particular, the four-point fermionic self-interaction has been left out.

In this paper, we initiate a careful study of the symmetries and interactions of the mass dimension one fermions. It is shown that although the theory violates Lorentz symmetry, the Lagrangian is invariant under global Lorentz transformations. The Lorentz violation encountered in the theory is quite subtle since they do not involve any deformations of the relativistic dispersion relation. They are instead encoded in the form of preferred direction which can in principle be deciphered from the differential cross-sections. A study on the effects of Lorentz violation is beyond the scope of this paper but it will be considered elsewhere. As for the interactions, our initial focus is on the scalar bosons. While this does not exhaust all the possibilities, it does provide us with sufficient information to determine the types of interactions that are allowed. Later, we also consider the local $U(1)$ interactions. 

In the course of our investigation, we impose two conditions. The interaction must be renormalisable and that the $S$-matrix must be unitary. The first condition is imposed by the mass dimensionality of the field operators and their counter-terms while the later requires us to compute the total cross-sections and examine their high-energy behaviour. We find to our surprise, that the fermionic self-interaction, which by virtue of its mass dimensionality should be renormalisable and unitary, is in fact unitary violating at high-energy. Therefore, it cannot be a fundamental interaction of the theory. An important consequence which follows is that it excludes all the four-point interactions since they cannot be renormalisable without the counter-term generated by the self-interaction. This includes the electromagnetic interactions generated by the standard local $U(1)$ gauge symmetry. In sec.~\ref{4}, we show that the fermions have no renormalisable or unitary local $U(1)$ invariant interactions  thus making them natural dark matter candidates.


The paper is organised as follows. Section~\ref{symmetries} studies the symmetries of the theory. Sections~\ref{physical} and~\ref{higher} study the interactions of the spin-half and higher-spin mass dimension one fermions respectively. We find, for fermions with spin $j>\frac{1}{2}$, they have no renormalisable or unitary interactions. For spin-half fermions, the theory only permits the Yukawa interaction.

\section{Symmetries}\label{symmetries}
The mass dimension one fermionic field $\Lambda(x)$ and its adjoint $\dual{\Lambda}(x)$ are constructed from a complete set of spin-half Elko 
\begin{equation}
\Lambda(x)=(2\pi)^{-3/2}\int\frac{d^{3}p}{\sqrt{2mE_{\mathbf{p}}}}\sum_{\alpha}[e^{-ip\cdot x}\xi(\p,\alpha)a(\p,\alpha)+e^{ip\cdot x}\zeta(\p,\alpha)b^{\dag}(\p,\alpha)],\label{eq:field}
\end{equation}
\begin{equation}
\dual{\Lambda}(x)=(2\pi)^{-3/2}\int\frac{d^{3}p}{\sqrt{2mE_{\mathbf{p}}}}\sum_{\alpha}[e^{ip\cdot x}\dual{\xi}(\p,\alpha)a^{\dag}(\p,\alpha)+e^{-ip\cdot x}\dual{\zeta}(\p,\alpha)b(\p,\alpha)].
\label{eq:adjoint}
\end{equation}
where the annihilation and creation operators satisfy the anti-commutation relations
\begin{equation}
\{a(\p',\alpha'),a^{\dag}(\p,\alpha)\}=
\{b(\p',\alpha'),b^{\dag}(\p,\alpha)\}=
\delta_{\alpha'\alpha}\delta^{3}(\p'-\p).
\end{equation}
One can also construct another set of fermionic field and adjoint where the particles and anti-particles are indistinguishable
\begin{equation}
\lambda(x)=\Lambda(x)\big\vert_{b=a},\hspace{0.5cm}
\dual{\lambda}(x)=\dual{\Lambda}(x)\big\vert_{b=a}
\end{equation}
but they will not be considered here. The definition and solutions of the spinors $\xi(\p,\alpha)$ and $\zeta(\p,\alpha)$ are given in app.~\ref{A}. As for the dual spinors $\dual{\xi}(\p,\alpha)$ and $\dual{\zeta}(\p,\alpha)$, their properties are discussed in sec.~\ref{spinor_dual}. The theory has many interesting properties, for a comprehensive review, please see~\cite{Ahluwalia:2013uxa}. Among the novel features, the most important ones are:
\begin{enumerate}
\item The field satisfies the Klein-Gordon but not the Dirac equation. 
\item The field is of mass dimension one and not three-half.
\item The existence of a preferred direction.
\end{enumerate}
These features are best characterized by the propagator
\begin{eqnarray}
S(x,y)
&=& \frac{i}{2}\int \frac{d^4 p}{(2\pi)^4} e^{-i p\cdot(x- y)} 
\left[ \frac{I + {\mathcal G}(\phi)}{p\cdot p - m^2 + i\epsilon}\right] \label{eq:prop}
\end{eqnarray}
where $\mathcal{G}(\phi)$ is an off-diagonal matrix
\begin{equation}
\mathcal{G}(\phi)=i\left(\begin{matrix}
0 & 0 & 0 & -e^{-i\phi} \\
0 & 0 & e^{i\phi} & 0 \\
0 & -e^{-i\phi} & 0 & 0\\
e^{i\phi} & 0 & 0 & 0 \end{matrix}\right).
\end{equation}
The angle $\phi$ is defined by the momentum $\p$
\begin{equation}
\p=|\p|(\sin\theta\cos\phi,\sin\theta\sin\phi,\cos\theta)\label{eq:spc}
\end{equation}
with $0\leq\theta\leq\pi$ and $0\leq\phi<2\pi$. In the propagator, the preferred direction is encoded in the $\mathcal{G}(\phi)$ matrix and its integral which is non-vanishing unless $\x-\y$ is aligned to the 3-axis. The propagator $S(x,y)$ is different from the one given in~\cite{Ahluwalia:2013uxa} by a factor of $\frac{1}{2}$. This is because here we have chosen a different normalisation for the spinors. As a result, the norms of the spinors and the spin-sums to be discussed will also differ from those in~\cite{Ahluwalia:2013uxa} by a factor of 2 and $\frac{1}{2}$ respectively. The reason for our choice is explained in app.~\ref{A}.


The theory is described by the Klein-Gordon Lagrangian despite the fact that the propagator is not a Green's function of the Klein-Gordon operator. This may initially seem to be undesirable, but in our opinion, it is an integral part of the theory that reflects the effect of the preferred direction and cannot be ignored. Additionally, our choice is supported by the fact that the Klein-Gordon Lagrangian yields a positive-definite free Hamiltonian~\cite{Lee:2012td}. Therefore, our focus should be on the physical consequences of the $\mathcal{G}(\phi)$ matrix and the preferred direction. It should be noted that
there has been attempt to define an operator in which the propagator is a Green's function but it requires a deformation where one takes $I+\mathcal{G}(\phi)$ to $I+\tau\mathcal{G}(\phi)$ with $\tau$ being a real number infinitesimally close to unity~\cite{Lee:2014opa}. However, it is unclear whether this procedure is mathematically well-defined.

Taking the Klein-Gordon Lagrangian as the starting point, we study the simplest physical processes involving mass dimension one fermions and real scalar bosons. These processes are easy to compute and provide valuable information on the structures of the theory. Based on the mass dimensionality of the field operators, we propose the following Lagrangian
\begin{eqnarray}
&&\mathscr{L}=\mathscr{L}_{0}+\mathscr{L}_{1},\\
&&\mathscr{L}_{0}=\partial^{\mu}\dual{\Lambda}\partial_{\mu}\Lambda-m_{\Lambda}^{2}\dual{\Lambda}\Lambda
+\frac{1}{2}(\partial^{\mu}\phi\partial_{\mu}\phi-m_{\phi}^{2}\phi^{2}),\\
&&\mathscr{L}_{1}=-g_{\phi}\dual{\Lambda}\Lambda\phi-\frac{g_{\phi^{2}}}{2}\dual{\Lambda}\Lambda\phi^{2}
-\frac{g_{\Lambda}}{2}(\dual{\Lambda}\Lambda)^{2}
-\frac{h_{\phi^{3}}}{3!}\phi^{3}-\frac{h_{\phi^{4}}}{4!}\phi^{4}. \label{eq:L1}
\end{eqnarray}
Later, we will also consider local $U(1)$ interactions. The remainder of this section focuses on the dual spinors, the field adjoint. Physical processes are studied in sec.~\ref{physical}.

\subsection{Dual spinors and adjoint}\label{spinor_dual}
An important property that distinguishes Elko from the Dirac spinors is that they have vanishing norms under the Dirac dual. Taking $\chi(\p,\alpha)$ to represent both the $\xi(\p,\alpha)$ and $\zeta(\p,\alpha)$ spinors, this means that
\begin{equation}
\overline{\chi}(\p,\alpha)\chi(\p,\alpha)=0
\end{equation} 
where $\overline{\chi}(\p,\alpha)=\chi^{\dag}(\p,\alpha)\eta$ and
\begin{equation}
\eta=\left(\begin{matrix}
O & I \\
I & O
\end{matrix}\right).
\end{equation}
In fact, by computing all its bilinear covariants, one finds that Elko is a flag-pole spinor of the Lounesto classification~\cite{daRocha:2005ti,Lounesto:2001zz}. 
The dual which yields non-vanishing orthonormal norms for Elko can be formally defined as~\cite{Ahluwalia:2013uxa,Speranca:2013hqa}
\begin{equation}
\dual{\chi}(\p,\alpha)=[\Xi(\p)\chi(\p,\alpha)]^{\dag}\eta\label{eq:dual}
\end{equation}
where
\begin{equation}
\Xi(\p)=\frac{1}{m}\mathcal{G}(\phi)\slashed{p}.
\end{equation}
The $\gamma^{\mu}$ matrices in the basis we are working with are
\begin{equation}
\gamma^{0}=\left(\begin{matrix}
O & I \\
I & O
\end{matrix}\right),\hspace{0.5cm}
\gamma^{i}=\left(\begin{matrix}
O & -\sigma^{i} \\
\sigma^{i} & O
\end{matrix}\right).
\end{equation}
The resulting Elko norms are given by
\begin{equation}
\dual{\xi}(\p,\alpha)\xi(\p,\alpha')
=-\dual{\zeta}(\p,\alpha)\zeta(\p,\alpha')=m\delta_{\alpha\alpha'}.
\end{equation}
As we shall see in sec.~\ref{physical}, an advantage of eq.~(\ref{eq:dual}) is that it simplifies the computation of the transition probability. This is possible because eq.~(\ref{eq:dual}) can be rewritten as
\begin{equation}
\dual{\chi}(\p,\alpha)=\overline{\chi}(\p,\alpha)\Xi(\p).\label{eq:dual2}
\end{equation}
But here it is more instructive to express the dual in its original form~\cite{Ahluwalia:2004ab} which can be obtained by an explicit evaluation of eq.~(\ref{eq:dual})
\begin{equation}
\dual{\chi}(\p,\alpha)=i(-1)^{1/2+\alpha}\overline{\chi}(\p,-\alpha). \label{eq:dual3}
\end{equation}
This means that $\dual{\chi}(\p,\alpha)$ has the same transformation properties as $\overline{\chi}(\p,\alpha)$. An important consequence that follows from eq.~(\ref{eq:dual3}) is  that the adjoint $\dual{\Lambda}(x)$ can be expanded in terms of $\overline{\chi}(\p,\alpha)$. Therefore, under a global transformation
\begin{equation}
\Lambda(x)\rightarrow\mathcal{D}(\mathcal{L})\Lambda(x)
\end{equation}
where $\mathcal{D}(\mathcal{L})$ is an element of the $(\frac{1}{2},0)\oplus(0,\frac{1}{2})$ representation of the Lorentz group, the field adjoint transforms as
\begin{equation}
\dual{\Lambda}(x)\rightarrow\dual{\Lambda}(x)\mathcal{D}^{-1}(\mathcal{L}).
\end{equation}
As a result, the inner-product $\dual{\Lambda}(x)\Lambda(x)$ and hence the Lagrangian $\mathscr{L}(x)$ are invariant under global Lorentz transformations despite the fact that the theory violates Lorentz symmetry.
\subsection{Symmetries of the spin-sums}
Lorentz violation of the theory is best characterized by the following Elko spin-sums which also determine the form of the propagator
\begin{equation}
\sum_{\alpha}\chi(\p,\alpha)\dual{\chi}(\p,\alpha)=\frac{m}{2}\left[\mathcal{G}(\phi)\pm I\right]
\label{eq:spin_sum}
\end{equation}
where the top and bottom signs apply to $\xi(\p,\alpha)$ and $\zeta(\p,\alpha)$ respectively. They are not covariant under arbitrary spinor transformation $\mathcal{D}(\mathcal{L})$. Our task here is to determine the global symmetries of the spin-sums.

Let $M$ be an arbitrary matrix that is not restricted to the $(\frac{1}{2},0)\oplus(0,\frac{1}{2})$ representation. The transformations on the spinors and its duals are
\begin{eqnarray}
\chi(\p,\sigma)&\rightarrow& M\chi(\p,\sigma),\\
\dual{\chi}(\p,\sigma)&\rightarrow&\dual{\chi}(\p,\sigma)(\eta M^{\dag}\eta).
\end{eqnarray}
Therefore, the spin-sums become
\begin{equation}
\frac{m}{2}[\mathcal{G}(\phi)\pm I]\rightarrow\frac{m}{2}
M[\mathcal{G}(\phi)\pm I](\eta M^{\dag}\eta).
\end{equation}
The demand of covariance or invariance requires the following conditions to be satisfied
\begin{eqnarray}
&&M(\eta M^{\dag}\eta)= I, \label{eq:c1}\\
&&M\mathcal{G}(\phi)(\eta M^{\dag}\eta)=\mathcal{G}(\phi+\bar{\phi}) \label{eq:c2}
\end{eqnarray}
where $\bar{\phi}$ is a constant. From eq.~(\ref{eq:c1}), we obtain
\begin{equation}
M\mathcal{G}(\phi)M^{-1}=\mathcal{G}(\phi+\bar{\phi})\label{eq:c3}
\end{equation}
where the most general solution is given by
\begin{equation}
M=\left(\begin{matrix}
a & 0 &  b & 0 \\
0 & e^{i\bar{\phi}}c & 0 & e^{i\bar{\phi}}d\\
d & 0 & c & 0 \\
0 & e^{i\bar{\phi}}b & 0 & e^{i\bar{\phi}}a \label{eq:m}
\end{matrix}\right).
\end{equation}
Substituting eq.~(\ref{eq:m}) into eq.~(\ref{eq:c1}) and imposing the condition
\begin{equation}
ac-bd=1,
\end{equation}
we obtain
\begin{eqnarray}
&& a=a^{*},\hspace{0.5cm} b=-b^{*},\\
&& c=c^{*},\hspace{0.5cm} d=-d^{*}
\end{eqnarray}
so that $a$ and $c$ are real while $b$ and $d$ are imaginary. We want the matrix $M$ to correspond to a global continuous space-time symmetry transformation. This means that $M$ must be an element of a Lie group so we must have $b=d=0$. Therefore,
\begin{equation}
M=\left(\begin{matrix}
a & 0 &  0 & 0 \\
0 & e^{i\bar{\phi}}a^{-1} & 0 & 0\\
0 & 0 & a^{-1} & 0 \\
0 & 0 & 0 & e^{i\bar{\phi}}a^{-1}
\end{matrix}\right).
\end{equation}
If we identify $M$ to be an element of the $(\frac{1}{2},0)\oplus(0,\frac{1}{2})$ representation, then the solutions which leave the spin-sums invariant and covariant are the boost and rotation about the 3-axis respectively in agreement with the results obtained in~\cite{Ahluwalia:2010zn}. The solutions for $a$ and $\bar{\phi}$ which give us boost and rotation are
\begin{equation}
M(a,\bar{\phi})=
\begin{cases}
\exp(iJ_{3}\bar{\phi}), & a=e^{i\bar{\phi}/2}\\
\exp(iK_{3}\varphi_{q}), & a=\sqrt{\frac{m}{E_{q}-q}},\bar{\phi}=0
\end{cases}
\end{equation}
where $\bar{\phi}$ is the angle of rotation and $\varphi_{q}$ is the rapidity parameter. It should be noted that the momentum $q$ is independent of the momentum ascribed to the spinors. Together, these transformations describe the global symmetry of the spin-sums and the propagator. They form the Abelian Lie group $SO(2)\times SO(1,1)$.

\section{Physical processes}\label{physical}

In this section, we study the physical processes involving mass dimension one fermions.  We start with the four-point fermionic self-interaction. Subsequently, we study their interactions with real scalar and vector bosons. Here we impose the condition that the interactions must satisfy unitarity.
For the $1+2\rightarrow1'+2'$ process in which we are considering, this means that at high-energy, the total cross-section must behave as $\sigma(12\rightarrow1'2')\sim\frac{1}{E^{2}}$ where $E$ is the centre of mass energy~\cite{Greiner:1993qp}. Contrary to the naive expectation of power-counting and renormalisation, the self-interaction is shown to violate unitarity at high-energy and therefore cannot be a fundamental interaction. This result has important consequences for the theory to be discussed in the following sections.

\subsection{Self-interaction}\label{self}

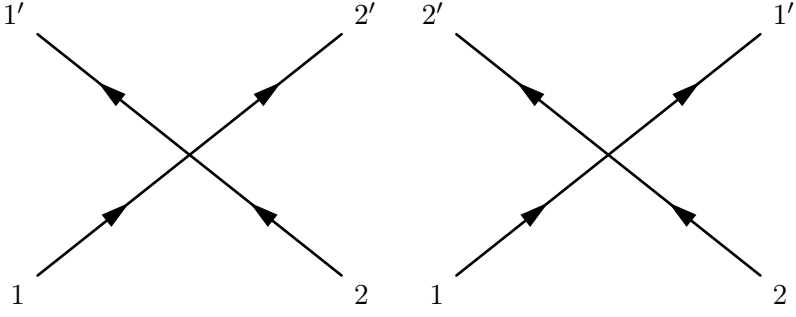
\begin{figure}
\begin{center}
\begin{fmffile}{1}
\begin{fmfgraph*}(40,40)
\fmftop{i1,i2}
\fmfbottom{o1,o2}
\fmf{fermion,tension=10}{o1,v1}
\fmf{fermion,tension=10}{o2,v1}
\fmf{fermion,tension=10}{v1,i1}
\fmf{fermion,tension=10}{v1,i2}
\fmflabel{1}{o1}
\fmflabel{2}{o2}
\fmflabel{$1'$}{i1}
\fmflabel{$2'$}{i2}
\end{fmfgraph*}
\end{fmffile}\hspace{1cm}
\begin{fmffile}{2}
\begin{fmfgraph*}(40,40)
\fmftop{i1,i2}
\fmfbottom{o1,o2}
\fmf{fermion,tension=10}{o1,v1}
\fmf{fermion,tension=10}{o2,v1}
\fmf{fermion,tension=10}{v1,i1}
\fmf{fermion,tension=10}{v1,i2}
\fmflabel{1}{o1}
\fmflabel{2}{o2}
\fmflabel{$2'$}{i1}
\fmflabel{$1'$}{i2}
\end{fmfgraph*}
\end{fmffile}
\end{center}
\caption{Fermionic self-interaction}\label{fig1}
\end{figure}

We consider the process $\Lambda_{1}+\Lambda_{2}\rightarrow\Lambda'_{1}+\Lambda'_{2}$ described by the interaction $\mathscr{H}_{\Lambda}=\frac{1}{2}g_{\Lambda}(\dual{\Lambda}\Lambda)^{2}$. At tree-level (fig.~\ref{fig1}), the $S$-matrix evaluates to
\begin{eqnarray}
S_{(\Lambda'_{1}\Lambda'_{2})(\Lambda_{1}\Lambda_{2})}&=&\frac{-ig_{\Lambda}\delta^{4}(p'_{1}+p'_{2}-p_{1}-p_{2})}{2(2\pi)^{2}\sqrt{16m^{4}_{\Lambda}E'_{1}E'_{2}E_{1}E_{2}}}[(\dual{\xi}'_{2}\xi_{2})(\dual{\xi}'_{1}\xi_{1})-(1\leftrightarrow2)-(1'\leftrightarrow2')\nonumber\\
&&\hspace{7cm}+(1\leftrightarrow2,1'\leftrightarrow2')]\nonumber\\
&=&\frac{-ig_{\Lambda}\delta^{4}(p'_{1}+p'_{2}-p_{1}-p_{2})}{(2\pi)^{2}\sqrt{16m^{4}_{\Lambda}E'_{1}E'_{2}E_{1}E_{2}}}[(\dual{\xi}'_{2}\xi_{2})(\dual{\xi}'_{1}\xi_{1})-(\dual{\xi}'_{1}\xi_{2})(\dual{\xi}'_{2}\xi_{1})]
\end{eqnarray}
where the sums takes into account of the fermionic statistics. Here we have adopted an abbreviated notation where $\xi_{i}=\xi(\p_{i},\alpha_{i})$ and $\xi'_{i}=\xi(\p'_{i},\alpha'_{i})$. Defining the $S$-matrix to be 
\begin{equation}
S_{\beta\alpha}=-2\pi i\delta^{4}(p_{\beta}-p_{\alpha})M_{\beta\alpha},
\end{equation}
the scattering amplitude is
\begin{equation}
M_{(\Lambda'_{1}\Lambda'_{2})(\Lambda_{1}\Lambda_{2})}=\frac{g_{\Lambda}}{(2\pi)^{3}\sqrt{16m^{4}_{\Lambda}E'_{1}E'_{2}E_{1}E_{2}}}[(\dual{\xi}'_{2}\xi_{2})(\dual{\xi}'_{1}\xi_{1})-(\dual{\xi}'_{1}\xi_{2})(\dual{\xi}'_{2}\xi_{1})].
\end{equation}
Instead of using eq.~(\ref{eq:dual}),  the adjoint $M^{\dag}_{(\Lambda'_{1}\Lambda'_{2})(\Lambda_{1}\Lambda_{2})}$ can be more easily computed using the formal definition given by eq.~(\ref{eq:dual2})
\begin{equation}
M^{\dag}_{(\Lambda'_{1}\Lambda'_{2})(\Lambda_{1}\Lambda_{2})}=\frac{g_{\Lambda}}{(2\pi)^{3}\sqrt{16m^{4}_{\Lambda}E'_{1}E'_{2}E_{1}E_{2}}}[(\overline{\xi}_{2}\Xi'_{2}\xi'_{2})(\overline{\xi}_{1}\Xi'_{1}\xi'_{1})-(\overline{\xi}_{2}\Xi'_{1}\xi'_{1})(\overline{\xi}_{1}\Xi'_{2}\xi'_{2})].
\end{equation}
 Therefore, the `spin-averaged' transition probability is proportional to
\begin{eqnarray}
\sum_{\mbox{\tiny{spins}}}|M_{(\Lambda'_{1}\Lambda'_{2})(\Lambda_{1}\Lambda_{2})}|^{2}&=&
\frac{g_{\Lambda}^{2}}{(2\pi)^{6}(16m^{4}_{\Lambda}E'_{1}E'_{2}E_{1}E_{2})}\nonumber\\
&&\times\sum_{\mbox{\tiny{spins}}}
[\mbox{tr}(\xi'_{2}\overline{\xi}'_{2}\xi_{2}\overline{\xi}_{2})
\mbox{tr}(\xi_{1}\overline{\xi}_{1}\xi'_{1}\overline{\xi}'_{1})
-\mbox{tr}(\xi'_{1}\overline{\xi}'_{1}\xi_{2}\overline{\xi}_{2}\xi'_{2}\overline{\xi}'_{2}
\xi_{1}\overline{\xi}_{1}) \nonumber\\
&&\hspace{0.5cm}-\mbox{tr}
(\xi'_{1}\overline{\xi}'_{1}\xi_{1}\overline{\xi}_{1}\xi'_{2}\overline{\xi}'_{2}
\xi_{2}\overline{\xi}_{2})
+\mbox{tr}(\xi'_{1}\overline{\xi}'_{1}\xi_{2}\overline{\xi}_{2})\mbox{tr}
(\xi'_{2}\overline{\xi}'_{2}\xi_{1}\overline{\xi}_{1})].\label{eq:self_prob}
\end{eqnarray}
In obtaining the above expression, we have used the fact that the spin-sums defined with respect to the Dirac dual are given by
\begin{equation}
\sum_{\alpha}\chi(\p,\alpha)\overline{\chi}(\p,\alpha)=\frac{1}{2}[I\pm\mathcal{G}(\phi)]\slashed{p}
\label{eq:Dirac_dual}
\end{equation}
which commutes with $\Xi(\p)$ since
\begin{equation}
[\slashed{p},\Xi(\p)]=[\mathcal{G}(\phi),\Xi(\p)]=O
\end{equation}
and that 
\begin{equation}
\Xi^{2}(\p)=I.
\end{equation}
According to our normalisations of the $S$-matrix and field operators, the differential cross-section in the centre of mass frame for a general $12\rightarrow1'2'$ process is given by~\cite[sec.~3.4]{Weinberg:1995mt}
\begin{equation}
\frac{d\sigma}{d\Omega_{\mbox{\tiny{CM}}}}(12\rightarrow1'2')=
\frac{(2\pi)^{4} |\p'_{1}| E'_{1}E'_{2}E_{1}E_{2}}{E^{2}|\p_{1}|}|M_{(1'2')(12)}|^{2}\label{eq:diff_cross_section}
\end{equation}
where $E=E_{1}+E_{2}$ and $d\Omega=\sin\theta'_{1}d\theta'_{1}d\phi'_{1}$. In our case, since all the external particles have the same mass, the differential cross-section simplifies to
\begin{equation}
\frac{d\sigma}{d\Omega_{\mbox{\tiny{CM}}}}(\Lambda_{1}\Lambda_{2}\rightarrow\Lambda'_{1}\Lambda'_{2})=
\frac{(2\pi)^{4}E^{2}}{16} |M_{(\Lambda'_{1}\Lambda'_{2})(\Lambda_{1}\Lambda_{2})}|^{2}.
\end{equation}
Therefore, the total spin-averaged cross-section is given by
\begin{equation}
\sigma_{\mbox{\tiny{avg}}}(\Lambda_{1}\Lambda_{2}\rightarrow\Lambda'_{1}\Lambda'_{2})=
\frac{(2\pi)^{4}E^{2}}{256}\int d\Omega\sum_{\mbox{\tiny{spins}}}|M_{(\Lambda'_{1}\Lambda'_{2})(\Lambda_{1}\Lambda_{2})}|^{2}.
\label{eq:total_cross}
\end{equation}
To simplify the calculation, we take the momenta of the incoming particles to be along the 3-axis
\begin{equation}
p_{1}=\left(\frac{E}{2},0,0,|\p|\right),\hspace{0.5cm} p_{2}=\left(\frac{E}{2},0,0,-|\p|\right)
\label{eq:momentum}
\end{equation}
Substituting eq.~(\ref{eq:self_prob}) into (\ref{eq:total_cross}), a direct evaluation yields
\begin{equation}
\sigma_{\mbox{\tiny{avg}}}(\Lambda_{1}\Lambda_{2}\rightarrow\Lambda'_{1}\Lambda'_{2})=
\frac{g_{\Lambda}^{2}}{6144\pi m^{4}_{\Lambda}E^{2}}
(7E^{4}-16m^{2}_{\Lambda}E^{2}+36m^{4}_{\Lambda}).
\end{equation}
 At high-energy, the cross-section behaves as $g^{2}_{\Lambda}(E^{2}/m^{4}_{\Lambda})$. Therefore, it will inevitably violate unitarity.
Demanding unitarity to be satisfied up to arbitrarily high-energy, the self-interaction must be excluded from the theory. 
 
 Here we can just consider the differential cross-section since it has the same high-energy behaviour as the total cross-section. Generally, unless there exists certain rotation symmetries such that the dominant energy contributions vanish upon integration, it is sufficient to consider the behaviour of the differential cross-section.  When we study physical processes for higher-spin fermions in sec.~\ref{higher}, it is assumed that no such symmetries exist.

The above result maybe surprising at first since the fermionic fields are of mass dimension one so one would naively expect the self-interaction to be renormalisable and unitary. Upon closer examination, it is not difficult to understand why this is not the case. The spin-sums that contribute to the differential cross-section is given by eq.~(\ref{eq:Dirac_dual}) instead of~(\ref{eq:spin_sum}). Had it been the later, eq.~(\ref{eq:total_cross}) would behave as $g^{2}_{\Lambda}/E^{2}_{\Lambda}$ and would be unitary. To obtain a unitary self-interaction, one possibility that has been explored was to redefine the transition probability such that the spin-sums that contribute to the cross-section is eq.~(\ref{eq:spin_sum})~\cite[sec.~4.3.1]{Lee:2013cwa}. However, the proposed definition is not mathematically well-defined and the resulting probability is not generally positive-definite. In app.~\ref{B}, we show that although the dual spinors for $\dual{\Lambda}(x)$ is not given by the Dirac dual, no contradictions arise when the particle annihilation and creation operators are related by Hermitian conjugation. Since the standard definition $|M_{\beta\alpha}|^{2}\equiv M^{\dag}_{\beta\alpha}M_{\beta\alpha}$ is always positive-definite, it remains the correct choice even though this means that the self-interaction cannot be a fundamental interaction. 

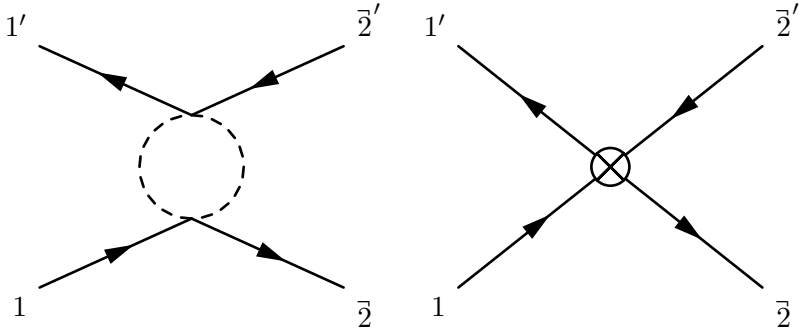
\begin{figure}
\begin{center}
\begin{fmffile}{3}
\begin{fmfgraph*}(40,40)
\fmftop{i1,i2}
\fmfbottom{o1,o2}
\fmf{fermion,tension=3}{o1,v1}
\fmf{fermion,tension=3}{v1,o2}
\fmf{fermion,tension=3}{v2,i1}
\fmf{fermion,tension=3}{i2,v2}
\fmf{dashes,left,tension=2}{v1,v2,v1}
\fmflabel{$1$}{o1}
\fmflabel{$\dual{2}$}{o2}
\fmflabel{$1'$}{i1}
\fmflabel{$\dual{2}'$}{i2}
\end{fmfgraph*}
\end{fmffile}\hspace{1cm}
\begin{fmffile}{foo}
\begin{fmfgraph*}(40,40)

\fmfcmd{%
vardef cross_bar (expr p, len, ang) =
((-len/2,0)--(len/2,0))
rotated (ang + angle direction length(p)/2 of p)
shifted point length(p)/2 of p
enddef;
style_def crossed expr p =
cdraw p;
ccutdraw cross_bar (p, 5mm, 45);
ccutdraw cross_bar (p, 5mm, -45)
enddef;}

\fmfcmd{
    path quadrant, q[], otimes;
    quadrant = (0, 0) -- (0.5, 0) & quartercircle & (0, 0.5) -- (0, 0);
    for i=1 upto 4: q[i] = quadrant rotated (45 + 90*i); endfor
    otimes = q[1] & q[2] & q[3] & q[4] -- cycle;
}
\fmfwizard

\fmftop{i1,i2}
\fmfbottom{o1,o2}

\fmf{fermion}{o1,v}
\fmf{fermion}{v,o2}
\fmf{fermion}{v,i1}
\fmf{fermion}{i2,v}

\fmfv{d.sh=otimes,d.f=empty}{v}
\fmflabel{$1$}{o1}
\fmflabel{$\dual{2}$}{o2}
\fmflabel{$1'$}{i1}
\fmflabel{$\dual{2}'$}{i2}
\end{fmfgraph*}
\end{fmffile}
\end{center}
\caption{A one-loop diagram with the counter-term required to cancel the divergence.}\label{fig2}
\end{figure}

\subsection{Scalar-interactions}

\begin{figure}
\begin{center}
\begin{fmffile}{4}
\begin{fmfgraph*}(40,40)
\fmftop{i1,i2}
\fmfbottom{o1,o2}
\fmf{fermion,tension=10}{o1,v1,v2,i2}
\fmf{dashes,tension=10}{o2,v1}
\fmf{dashes,tension=10}{v2,i1}
\fmflabel{1}{o1}
\fmflabel{2}{o2}
\fmflabel{$2'$}{i1}
\fmflabel{$1'$}{i2}
\end{fmfgraph*}
\end{fmffile}\hspace{1cm}
\begin{fmffile}{5}
\begin{fmfgraph*}(40,40)
\fmftop{i1,i2}
\fmfbottom{o1,o2}
\fmf{fermion,tension=10}{o1,v1,v2,i2}
\fmf{dashes,tension=10}{o2,v2}
\fmf{dashes,tension=10}{v1,i1}
\fmflabel{1}{o1}
\fmflabel{2}{o2}
\fmflabel{$2'$}{i1}
\fmflabel{$1'$}{i2}
\end{fmfgraph*}
\end{fmffile}
\end{center}
\caption{Fermion-scalar scattering}\label{fig3}
\end{figure}
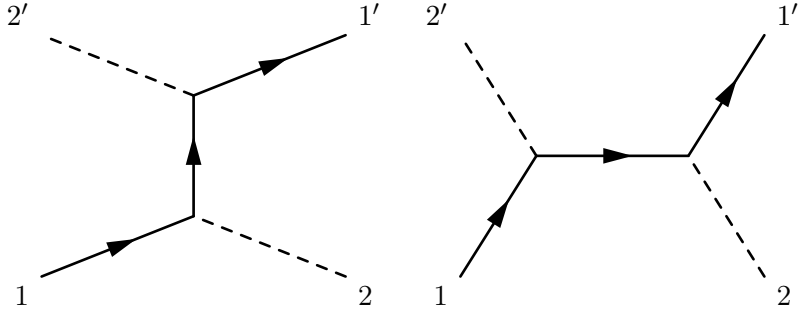

The exclusion of the self-interaction has important consequences for the theory. For the scalar interactions, it means that $\mathscr{H}_{\phi^{2}}=\frac{1}{2}g_{\phi^{2}}(\dual{\Lambda}\Lambda)\phi^{2}$ is ruled out. This is because at order $O(g_{\phi^{2}}^{2})$, the process $\Lambda_{1}+\dual{\Lambda}_{2}\rightarrow\Lambda'_{1}+\dual{\Lambda}'_{2}$ contains a scalar loop where the divergence can only be cancelled by the counter-term generated by $\mathscr{H}_{\Lambda}$~(fig.~\ref{fig2}). Therefore, without the self-interaction, $\mathscr{H}_{\phi^{2}}$ is non-renormalisable. The only scalar interaction that is renormalisable and unitary is the Yukawa interaction
\begin{equation}
\mathscr{L}_{\phi}=-g_{\phi}\dual{\Lambda}\Lambda\phi
\end{equation}
from which we shall consider a simple tree-level process $\Lambda_{1}+\phi_{2}\rightarrow\Lambda'_{1}+\phi_{2}'$ (fig.~\ref{fig3}). In this case, the amplitude is given by
\begin{equation}
M_{(\Lambda'_{1}\phi'_{2})(\Lambda_{1}\phi_{2})}=
\frac{g^{2}_{\phi}}{2(2\pi)^{3}\sqrt{16m^{2}_{\Lambda}E_{1}E_{2}E'_{1}E'_{2}}}\dual{\xi}'_{1}
\left[\frac{I+\mathcal{G}(\phi_{\mathbf{s}})}{s-m^{2}_{\Lambda}}
+\frac{I+\mathcal{G}(\phi_{\mathbf{u}})}{u-m^{2}_{\Lambda}}\right]\xi_{1} \label{eq:scalar}
\end{equation}
where $s=(p_{1}+p_{2})^{2}$ and $u=(p_{1}-p'_{2})^{2}$ are the Mandelstam variables. We shall leave computation of the cross-section associated with $M_{(\Lambda'_{1}\phi'_{2})(\Lambda_{1}\phi_{2})}$ to a later publication that is entirely devoted to the study of the scalar interaction. Here, useful information can already be extracted by examining the amplitude. Because of the $\mathcal{G}(\phi)$ matrix, the amplitude is only invariant under boost and rotation about the 3-axis. The only case where it becomes approximately Lorentz-invariant is when
the scalar boson is initially at rest with $\p_{2}=\0$ and $E_{1}\ll m_{\phi}$ so that $\p'_{2}\approx\0$. An interesting point worthy of note is that it is possible to introduce a contact interaction to cancel the non-covariant $\mathcal{G}(\phi)$ matrix thus giving us a Lorentz-invariant amplitude. But this is not physical since this term would be a function of two different space-time points. Therefore, it is non-local and does not have a well-defined unitary evolution in time.

\subsection{Local $U(1)$ interactions}\label{4}

\begin{figure}
\begin{center}
\begin{fmffile}{8}
\begin{fmfgraph*}(40,40)
\fmftop{i1,i2}
\fmfbottom{o1,o2}
\fmf{fermion}{o1,v1}
\fmf{fermion}{o2,v2}
\fmf{fermion}{v2,i2}
\fmf{fermion}{v1,i1}
\fmf{photon}{v1,v2}
\fmflabel{1}{o1}
\fmflabel{2}{o2}
\fmflabel{$1'$}{i1}
\fmflabel{$2'$}{i2}
\end{fmfgraph*}\hspace{1cm}
\end{fmffile}
\begin{fmffile}{9}
\begin{fmfgraph*}(40,40)
\fmftop{i1,i2}
\fmfbottom{o1,o2}
\fmf{fermion}{o1,v1}
\fmf{fermion}{o2,v2}
\fmf{fermion}{v2,i2}
\fmf{fermion}{v1,i1}
\fmf{photon}{v1,v2}
\fmflabel{1}{o1}
\fmflabel{2}{o2}
\fmflabel{$2'$}{i1}
\fmflabel{$1'$}{i2}
\end{fmfgraph*}
\end{fmffile}
\end{center}
\caption{Fermion scattering mediated by a vector boson.}\label{fig4}
\end{figure}
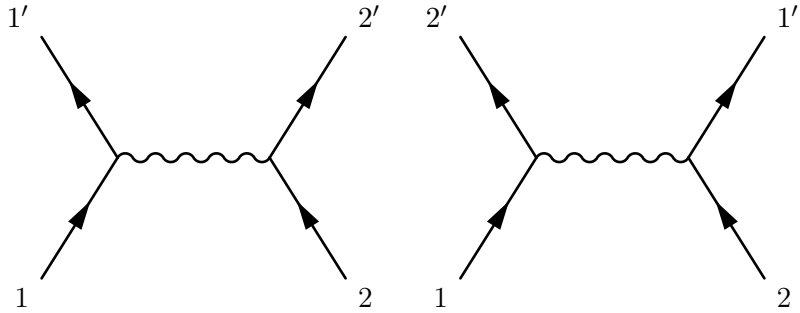

The demand of renormalisability and unitarity impose strong constraints on the possible interactions for the mass dimension one fermions. The exclusion of $\mathscr{H}_{\Lambda}$ has consequences beyond the scalar interactions under consideration. Repeating the argument in the previous section, it follows that all possible four-point interactions must be excluded on the grounds of renormalisability since at one-loop, they all require $\mathscr{H}_{\Lambda}$ to generate the counter-term to cancel the divergences. An important  interaction that is excluded is the the local $U(1)$ interaction. This of course, includes the electromagnetic interaction. But in general, the vector potential $a^{\mu}(x)$ associated with the $U(1)$ symmetry does not have to be identified with the photon, it could represent other new vector bosons from physics beyond the SM. But here the physical interpretation of $a^{\mu}(x)$ is not important. The important point is that such a Lagrangian takes the form $D^{\mu}\dual{\Lambda}D_{\mu}\Lambda$ where $D_{\mu}=\partial_{\mu}-ig_{a}a_{\mu}$. Although this term is invariant under local $U(1)$ transformation, but because it contains a four-point interaction of the form $-g_{a}^{2}\dual{\Lambda}\Lambda a^{\mu}a_{\mu}$, it is not renormalisable. 
But this does not exclude all possible local $U(1)$ interactions, another possibility is
\begin{equation}
\mathscr{L}_{f}=-g_{f}\dual{\Lambda}[\gamma^{\mu},\gamma^{\nu}]\Lambda f_{\mu\nu}\label{eq:f}
\end{equation}
where $g_{f}$ is the coupling constant and $f_{\mu\nu}=\partial_{\mu}a_{\nu}-\partial_{\nu}a_{\mu}$ is the field strength tensor. 
To see whether $\mathscr{L}_{f}$ gives us a unitary interaction, we compute the cross-section for fermion-fermion scattering mediated by a vector boson (fig.~\ref{fig4}). The amplitude $M_{(\Lambda'_{1}\Lambda'_{2})(\Lambda_{1}\Lambda_{2})}$ is given by
\begin{eqnarray}
M_{(\Lambda'_{1}\Lambda'_{2})(\Lambda_{1}\Lambda_{2})}&=&\frac{-4ig^{2}_{f}}{(2\pi)^{3}(16m^{4}_{\Lambda}E'_{1}E'_{2}E_{1}E_{2})^{1/2}}
\Bigg\{\left(\frac{1}{q^{2}}\right)\left[\bar{\xi}'_{1}\Xi'_{1}(iS^{\mu})\xi_{1}\right]
\left[\bar{\xi}'_{2}\Xi'_{2}(iS_{\mu})\xi_{2}\right]\nonumber\\
&&\hspace{4.5cm}-\left(\frac{1}{r^{2}}\right)\left[\bar{\xi}'_{2}\Xi'_{2}(iT^{\mu})\xi_{1}\right]
\left[\bar{\xi}'_{1}\Xi'_{1}(iT_{\mu})\xi_{2}\right]\Bigg\} \nonumber\\
\end{eqnarray}
where
\begin{equation}
q=p_{1}-p'_{1},\hspace{0.5cm}
r=p_{1}-p'_{2}
\end{equation}
and
\begin{equation}
S^{\nu}=q_{\mu}[\gamma^{\mu},\gamma^{\nu}],\hspace{0.5cm}
T^{\nu}=r_{\mu}[\gamma^{\mu},\gamma^{\nu}].
\end{equation}
The Hermitian conjugate of $M_{(\Lambda'_{1}\Lambda'_{2})(\Lambda_{1}\Lambda_{2})}$ is
\begin{eqnarray}
M^{\dag}_{(\Lambda'_{1}\Lambda'_{2})(\Lambda_{1}\Lambda_{2})}&=&\frac{4ig^{2}_{f}}{(2\pi)^{3}(16m^{4}_{\Lambda}E'_{1}E'_{2}E_{1}E_{2})^{1/2}}
\Bigg\{\left(\frac{1}{q^{2}}\right)\left[\bar{\xi}_{1}(iS^{\mu})\Xi'_{1}\xi'_{1}\right]
\left[\bar{\xi}_{2}(iS_{\mu})\Xi'_{2}\xi'_{2}\right]\nonumber\\
&&\hspace{4.5cm}-\left(\frac{1}{r^{2}}\right)\left[\bar{\xi}_{1}(iT^{\mu})\Xi'_{2}\xi'_{2}\right]
\left[\bar{\xi}_{2}(iT_{\mu})\Xi'_{1}\xi'_{1}\right]\Bigg\}\nonumber\\ \label{eq:Md}
\end{eqnarray}
where we have used the identity $\eta(S^{\mu})^{\dag}\eta=S^{\mu}$ and 
$\eta(T^{\mu})^{\dag}\eta=T^{\mu}$.
Therefore, the `spin-averaged' transition probability is proportional to
\begin{eqnarray}
\sum_{\mbox{\tiny{spins}}}|M_{(\Lambda'_{1}\Lambda'_{2})(\Lambda_{1}\Lambda_{2})}|^{2}&=&
\frac{g^{4}_{f}}{(2\pi)^{6}(m^{4}_{\Lambda}E'_{1}E'_{2}E_{1}E_{2})}\nonumber\\
&&\times\sum_{\mbox{\tiny{spins}}}
\Bigg[
\left(\frac{1}{q^{4}}\right)
\mbox{tr}(\xi_{1}\bar{\xi}_{1}S^{\mu}\xi'_{1}\bar{\xi}'_{1}S^{\nu})
\mbox{tr}(\xi_{2}\bar{\xi}_{2}S_{\mu}\xi'_{2}\bar{\xi}'_{2}S_{\nu})\nonumber\\
&&\hspace{1cm}-\left(\frac{1}{q^{2}r^{2}}\right)
\mbox{tr}(\xi_{1}\bar{\xi}_{1}T^{\mu}\xi'_{2}\bar{\xi}'_{2}S^{\nu}
\xi_{2}\bar{\xi}_{2}T_{\mu}\xi'_{1}\bar{\xi}'_{1}S_{\nu})\nonumber\\
&&\hspace{1cm}-\left(\frac{1}{q^{2}r^{2}}\right)
\mbox{tr}(\xi_{1}\bar{\xi}_{1}S^{\mu}\xi'_{1}\bar{\xi}'_{1}T^{\nu}
\xi_{2}\bar{\xi}_{2}S_{\mu}\xi'_{2}\bar{\xi}'_{2}T_{\nu})\nonumber\\
&&\hspace{1cm}+\left(\frac{1}{r^{4}}\right)
\mbox{tr}(\xi_{1}\bar{\xi}_{1}T^{\mu}\xi'_{2}\bar{\xi}'_{2}T^{\nu})
\mbox{tr}(\xi_{2}\bar{\xi}_{2}T_{\mu}\xi'_{1}\bar{\xi}'_{1}T_{\nu})
\Bigg].
\end{eqnarray}
Taking the momenta of the incoming particles to be given by eq.~(\ref{eq:momentum}) and by evaluating the individual traces and performing the angular integral according to eq.~(\ref{eq:total_cross}), the total spin-averaged cross-section is given by
\begin{equation}
\sigma_{\mbox{\tiny{avg}}}(\Lambda_{1}\Lambda_{2}\rightarrow \Lambda_{1}'\Lambda_{2}')=
\frac{g^{4}_{f}}{2m^{4}_{\Lambda}E^{2}}(3E^{4}-11m^{2}_{\Lambda}E^{2}+14m^{4}_{\Lambda}).
\end{equation}
At high-energy, the cross-section behaves as $g^{4}_{f}(E^{2}/m^{4}_{\Lambda})$. Therefore, like the self-interaction considered in sec.~\ref{self}, it will inevitably violate unitarity.

By the demand of renormalisability and unitarity, we see that the mass dimension one fermions cannot have local $U(1)$ invariant interactions. It follows that the fermions also cannot have non-Abelian interactions since these interactions, modulo the non-trivial Lie algebraic generators, take the same form as their Abelian counterparts so they are either non-renormalisable or unitary violating. The lack of local gauge-invariant interactions make the mass dimension one fermions natural dark matter candidates. Apart from gravity, their only interaction with the SM sector is through the Higgs boson via the Yukawa interaction.

\section{Higher-spin fermions} \label{higher}	
\begin{figure}
\begin{center}
\begin{fmffile}{6}
\begin{fmfgraph*}(40,40)
\fmftop{i1,i2}
\fmfbottom{o1,o2}
\fmf{fermion}{o1,v1}
\fmf{fermion}{o2,v2}
\fmf{fermion}{v2,i2}
\fmf{fermion}{v1,i1}
\fmf{dashes}{v1,v2}
\fmflabel{1}{o1}
\fmflabel{2}{o2}
\fmflabel{$1'$}{i1}
\fmflabel{$2'$}{i2}
\end{fmfgraph*}\hspace{1cm}
\end{fmffile}
\begin{fmffile}{7}
\begin{fmfgraph*}(40,40)
\fmftop{i1,i2}
\fmfbottom{o1,o2}
\fmf{fermion}{o1,v1}
\fmf{fermion}{o2,v2}
\fmf{fermion}{v2,i2}
\fmf{fermion}{v1,i1}
\fmf{dashes}{v1,v2}
\fmflabel{1}{o1}
\fmflabel{2}{o2}
\fmflabel{$2'$}{i1}
\fmflabel{$1'$}{i2}
\end{fmfgraph*}
\end{fmffile}
\end{center}
\caption{Fermion scattering mediated by a real scalar boson.}\label{fig5}
\end{figure}
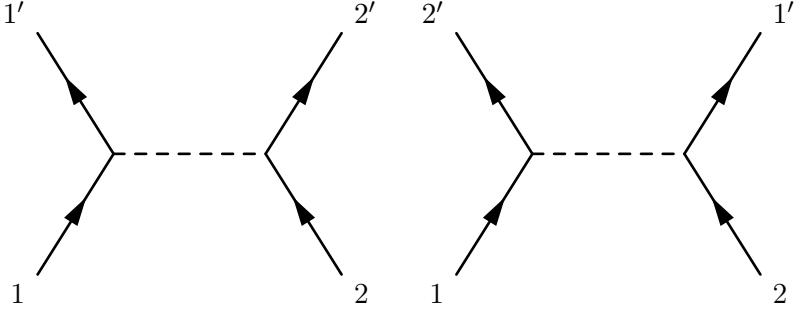

In this section, we study the interactions of the higher-spin fermionic fields constructed in~\cite{Lee:2012td}. These fields and their adjoints take the same form as those given in eqs.~(\ref{eq:field}) and (\ref{eq:adjoint}). The differences being that $\xi(\p,\alpha)$, $\zeta(\p,\alpha)$ and their duals are now replaced by their higher-spin generalisations. Independent of the spin, they are still of mass dimension one as can be seen from the propagator
\begin{eqnarray}
S^{(j)}(x,y)
&=& \frac{i}{2}\int \frac{d^4 p}{(2\pi)^4} e^{-i p\cdot(x- y)} 
\left[ \frac{I + {\mathcal G}^{(j)}(\phi)}{p\cdot p - m^2 + i\epsilon}\right]
\end{eqnarray}
where the matrix $\mathcal{G}^{(j)}(\phi)$ is given by
\begin{equation}
\mathcal{G}^{(j)}(\phi)=\left(\begin{matrix}
O & G \\
G& O
\end{matrix}\right),\hspace{0.5cm}
G_{\ell m}=(-1)^{j+\ell}e^{-2i\ell\phi}\delta_{\ell,-m}
\end{equation}
with $\ell,m=-j,\cdots,j$.

By virtue of the mass dimensionality of the fermionic fields, the naive interactions we would have proposed would be identical eq.~(\ref{eq:L1}). Using results obtained from sec.~\ref{physical}, the relevant spin-sums that contribute to the spin-averaged differential cross-sections are
\begin{equation}
\sum_{\alpha}\xi^{(j)}(\p,\alpha)\overline{\xi}^{(j)}(\p,\alpha)
=\frac{1}{2m^{2j-1}}[I+\mathcal{G}^{(j)}(\phi)]\gamma^{\mu_{1}\cdots\mu_{2j}}p_{\mu_{1}}\cdots p_{\mu_{2j}},\label{eq:spin_sum_j}
\end{equation}
\begin{equation}
\sum_{\alpha}\zeta^{(j)}(\p,\alpha)\overline{\zeta}^{(j)}(\p,\alpha)
=\frac{1}{2m^{2j-1}}[I-\mathcal{G}^{(j)}(\phi)]\gamma^{\mu_{1}\cdots\mu_{2j}}p_{\mu_{1}}\cdots p_{\mu_{2j}}
\end{equation}
where $\gamma^{\mu_{1}\cdots\mu_{2j}}$ is a symmetric traceless tensor of rank-$2j$.
Since their high-energy behaviour is of the order $E^{2j}_{\mathbf{p}}$, unitarity excludes the fermionic self-interaction for all spin. It follows that $\mathscr{H}^{(j)}_{\phi^{2}}$ and other four-point interactions are not renormalisable. We now show that $\mathscr{H}^{(j)}_{\phi}$ is also excluded by unitarity for fermionic fields with spin $j>\frac{1}{2}$.

For $\mathscr{H}^{(j)}_{\phi}$, let us consider the process of $\Lambda_{1}+\Lambda_{2}\rightarrow\Lambda'_{1}+\Lambda'_{2}$ mediated by a real scalar boson (fig.~\ref{fig5}). Since we have four external fermions, modulo the contribution from the scalar propagator, the spin-averaged transition  probability has the same form as eq.~(\ref{eq:self_prob}). Take into account of the high-energy behaviour of the spin-sum given by eq.~(\ref{eq:spin_sum_j})  and the scalar propagator, we get
\begin{equation}
\sum_{\mbox{\tiny{spins}}}|M_{(\Lambda'_{1}\Lambda'_{2})(\Lambda_{1}\Lambda_{2})}|^{2}\sim
g^{4}_{\phi}\left(\frac{1}{m^{4}_{\Lambda}E^{4}}\right)\left(\frac{1}{E^{4}}\right)
\left(\frac{E^{8j}}{m^{8j-4}_{\Lambda}}\right).
\end{equation}
The first term in the bracket comes from the normalisation of the fields while the second and third term come from the behaviour of the scalar propagator and the traces of the spin-sums. Therefore, we obtain
\begin{equation}
\frac{d\sigma_{\mbox{\tiny{avg}}}}{d\Omega_{\mbox{\tiny{CM}}}}(\Lambda_{1}\Lambda_{2}\rightarrow
\Lambda'_{1}\Lambda'_{2})
\sim g^{4}_{\phi}\left(\frac{E^{8j-6}}{m^{8j}_{\Lambda}}\right).
\end{equation}
Since this is unbounded for $j>\frac{1}{2}$, we see that unitarity is violated for higher-spin fermions  at high-energy. Because the high-energy behaviour of the spin-sum is of the order $E^{2j}_{\mathbf{p}}$, the cross-section associated with the local $U(1)$ interaction will also grow as positive powers of energy and violate unitarity. Therefore, despite the fields of mass dimension one, when $j>\frac{1}{2}$, they do not have any renormalisable or unitary interactions.
%
%

\section{Conclusions}

In this paper, we have studied the symmetries and interactions of the mass dimension one fermionic fields. A careful analysis of the dual spinors and the field adjoint shows that even though the theory violates Lorentz symmetry, the Lagrangian is in fact invariant under global Lorentz transformations. 
As for the particle interactions, we find that contrary to the naive power-counting argument, for fermionic fields with spin $j>\frac{1}{2}$, the demand of renormalisability and unitarity exclude all possible fundamental interactions. For the spin-half fermionic fields, the only allowed interaction is the Yukawa interaction thus making them natural dark matter candidates. Therefore, the full interacting Lagrangian is simply
\begin{equation}
\mathscr{L}_{1}=-g_{\phi}\dual{\Lambda}\Lambda\phi
-\frac{h_{\phi^{3}}}{3!}\phi^{3}-\frac{h_{\phi^{4}}}{4!}\phi^{4}.
\end{equation}

There are many important processes to be studied. A particular important class of processes are ones that involve fermionic loop corrections to the scalar propagator which contribute to $\delta m_{\phi}^{2}$. In the context of the SM, this will determine how the mass dimension one fermions contribute to the radiative corrections of the Higgs mass. Because the propagator is not Lorentz-covariant, its contribution to $\delta m_{\phi}^{2}$ would contain information on Lorentz violation in the form of a preferred direction. Therefore, apart from the direct searches at particle accelerators and constraints obtained from astrophysics and cosmology, a careful investigation of the radiative corrections to the Higgs mass is another promising avenue to detect the signatures of the mass dimension one fermions.
%

%

\acknowledgments

This research is supported by the CNPq grant 313285/2013-6. I would like to thank the generous hospitality offered by IUCAA where part of this work was completed.

\appendix

\section{Spin-half Elko}\label{A}
Elko is a complete set of four-component spinors constructed from the spinor
\begin{equation}
\chi(\p,\alpha)=\left[
\begin{matrix}
\vartheta\Theta\phi^{*}(\p,\alpha) \\
\phi(\p,\alpha)
\end{matrix}\right]
\end{equation}
where $\Theta=-i\sigma_{2}$ and $\phi(\p,\alpha)$ is a left-handed Weyl spinor. Elko is defined to be the eigenspinors of the charge conjugation operator
\begin{equation}
\mathcal{C}=\left(\begin{matrix}
O & i\Theta \\
-i\Theta & O
\end{matrix}\right)K
\end{equation}
where $K$ is an anti-unitary operator that complex conjugates all functions to its right. The spinor $\chi(\p,\alpha)$ becomes Elko with the following choice of phases
\begin{equation}
\mathcal{C}\chi(\p,\sigma)\vert_{\vartheta=\pm i}=\pm\chi(\p,\sigma)\vert_{\vartheta=\pm i}.
\end{equation}
Here, we define the spinors in the helicity basis. The left-handed Weyl spinors at rest are taken to be
\begin{equation}
\phi(\e,{\textstyle{\frac{1}{2}}})=\sqrt{\frac{m}{2}}
\left[\begin{matrix}
\cos(\theta/2)e^{-i\phi/2}\\
\sin(\theta/2)e^{i\phi/2} \end{matrix}\right],
\end{equation}
\begin{equation}
\phi(\e,-{\textstyle{\frac{1}{2}}})=\sqrt{\frac{m}{2}}
\left[\begin{matrix}
-\sin(\theta/2)e^{-i\phi/2} \\
\cos(\theta/2)e^{i\phi/2}\end{matrix}\right]
\end{equation}
where $\e=\lim_{|\mathbf{p}|\rightarrow\mathbf{0}}\hat{\p}$. The normalisations of $\phi(\e,\alpha)$ differ from those in~\cite{Ahluwalia:2013uxa} by a factor of $\frac{1}{\sqrt{2}}$. The normalisation is chosen such that the energy of a free particle state is $E_{\mathbf{p}}$ instead of $2E_{\mathbf{p}}$~\cite{Lee:2012td}.  The spinors at rest and at arbitrary momentum are related by
\begin{equation}
\phi(\p,\alpha)=\exp\left(-\frac{1}{2}\s\cdot\bv\right)\phi(\e,\alpha)
\end{equation} 
where $\bv=\varphi\hat{\p}$ is the rapidity parameter defined as
\begin{equation}
\cosh\varphi=\frac{E_{\mathbf{p}}}{m},\hspace{0.5cm}
\sinh\varphi=\frac{|\p|}{m}.
\end{equation} 
Using the identity $\Theta\s\Theta^{-1}=-\s^{*}$, we obtain
\begin{equation}
\chi(\p,\alpha)=\left[\begin{matrix}
\exp\left(\frac{1}{2}\s\cdot\bv\right) & O \\
O & \exp\left(-\frac{1}{2}\s\cdot\bv\right)
\end{matrix}\right]\chi(\e,\alpha).
\end{equation}
The solutions of Elko at rest are given by~\cite{Ahluwalia:2008xi,Ahluwalia:2009rh}
\begin{eqnarray}
\xi(\e,+\textstyle{\frac{1}{2}})&=&+\chi(\e,+\textstyle{\frac{1}{2}})\vert_{\vartheta=+i},\\
\xi(\e,-\textstyle{\frac{1}{2}})&=&+\chi(\e,-\textstyle{\frac{1}{2}})\vert_{\vartheta=+i},\\
\zeta(\e,+\textstyle{\frac{1}{2}})&=&+\chi(\e,-\textstyle{\frac{1}{2}})\vert_{\vartheta=-i},\\
\zeta(\e,-\textstyle{\frac{1}{2}})&=&-\chi(\e,+\textstyle{\frac{1}{2}})\vert_{\vartheta=-i}.
\end{eqnarray}

\section{The adjoint} \label{B}

Suppose that the mass dimension one fermionic field and its adjoint are given by
\begin{equation}
\Lambda(x)=(2\pi)^{-3/2}\int\frac{d^{3}p}{\sqrt{2mE_{\mathbf{p}}}}\sum_{\alpha}[e^{-ip\cdot x}\xi(\p,\alpha)a(\p,\alpha)+e^{ip\cdot x}\zeta(\p,\alpha)b^{\ddag}(\p,\alpha)],
\end{equation}
\begin{equation}
\dual{\Lambda}(x)=(2\pi)^{-3/2}\int\frac{d^{3}p}{\sqrt{2mE_{\mathbf{p}}}}\sum_{\alpha}[e^{ip\cdot x}\dual{\xi}(\p,\alpha)a^{\ddag}(\p,\alpha)+e^{-ip\cdot x}\dual{\zeta}(\p,\alpha)b(\p,\alpha)]
\end{equation}
where a new operator $\ddag$ has been introduced in place of the Hermitian conjugation $\dag$ for the creation operators. Here we show that although the spinors have a dual that differs from the standard Dirac dual, no contradictions arise from the identification $\ddag= \dag$.

First, we derive the expressions for $a(\p,\alpha)$ and $a^{\ddag}(\p,\alpha)$. This is achieved by performing Fourier inversion on $\Lambda(x)$ and $\dual{\Lambda}(x)$
\begin{equation}
a^{\ddag}(\p,\alpha)=(2\pi)^{-3/2}\int d^{3}x e^{-ip\cdot x}
\sqrt{\frac{E_{\mathbf{p}}}{8m}}
\left[\dual{\Lambda}(x)-\frac{i}{E_{\mathbf{p}}}\frac{\partial\dual{\Lambda}}{\partial t}(x)\right]\xi(\p,\alpha).\label{eq:addag}
\end{equation}
\begin{equation}
a(\p,\alpha)=(2\pi)^{-3/2}\int d^{3}x e^{ip\cdot x}\sqrt{\frac{E_{\mathbf{p}}}{8m}}
\dual{\xi}(\p,\alpha)\left[\Lambda(x)+\frac{i}{E_{\mathbf{p}}}\frac{\partial\Lambda}{\partial t}(x)\right].
\label{eq:a1}
\end{equation}
Perform similar calculations for $b(\p,\alpha)$ and $b^{\ddag}(\p,\alpha)$, we get
\begin{equation}
b(\p,\alpha)=(2\pi)^{-3/2}\int d^{3}x e^{ip\cdot x}
\sqrt{\frac{E_{\mathbf{p}}}{8m}}
\left[-\dual{\Lambda}(x)-\frac{i}{E_{\mathbf{p}}}\frac{\partial\dual{\Lambda}}{\partial t}(x)\right]\zeta(\p,\alpha),
\end{equation}
\begin{equation}
b^{\ddag}(\p,\alpha)=(2\pi)^{-3/2}\int d^{3}x e^{-ip\cdot x}\sqrt{\frac{E_{\mathbf{p}}}{8m}}
\dual{\zeta}(\p,\alpha)\left[-\Lambda(x)+\frac{i}{E_{\mathbf{p}}}\frac{\partial\Lambda}{\partial t}(x)\right].\label{eq:bddag}
\end{equation}

An important property of theory is that the anti-commutator between the field and its adjoint vanishes at space-like separation
\begin{equation}
\{\Lambda(t,\x),\dual{\Lambda}(t,\y)\}=O.
\end{equation}
This result requires the annihilation and creation operators to satisfy the canonical algebra
\begin{eqnarray}
\{a(\p',\alpha'),a^{\ddag}(\p,\alpha)\}&=&\delta_{\alpha'\alpha}\delta^{3}(\p'-\p),\\
\{b(\p',\alpha'),b^{\ddag}(\p,\alpha)\}&=&\delta_{\alpha'\alpha}\delta^{3}(\p'-\p).
\end{eqnarray}

The field adjoint $\dual{\Lambda}(x)$ is different from $\overline{\Lambda}(x)$ so that $[\dual{\Lambda}(x)\chi(\p,\alpha)]^{\dag}\neq[\dual{\chi}(\p,\alpha)\Lambda(x)]$. It suggests that $\ddag$ cannot be identified as $\dag$ but this is not true. Such an identification is in fact possible. To see this, we prove that no contradictions arise from the following relations
\begin{eqnarray}
a^{\ddag}(\p,\alpha)&=& a^{\dag}(\p,\alpha),\label{eq:id1}\\ 
b^{\ddag}(\p,\alpha)&=& b^{\dag}(\p,\alpha). \label{eq:id2}
\end{eqnarray}
From eqs.~(\ref{eq:id1}) and (\ref{eq:id2}), the creation operators given in eqs.~(\ref{eq:addag}) and (\ref{eq:bddag}) become
\begin{equation}
a^{\dag}(\p,\alpha)=(2\pi)^{-3/2}\int d^{3}x e^{-ip\cdot x}
\sqrt{\frac{E_{\mathbf{p}}}{8m}}
\left[\dual{\Lambda}(x)-\frac{i}{E_{\mathbf{p}}}\frac{\partial\dual{\Lambda}}{\partial t}(x)\right]\xi(\p,\alpha),
\end{equation}
\begin{equation}
b^{\dag}(\p,\alpha)=(2\pi)^{-3/2}\int d^{3}x e^{-ip\cdot x}\sqrt{\frac{E_{\mathbf{p}}}{8m}}
\dual{\zeta}(\p,\alpha)\left[-\Lambda(x)+\frac{i}{E_{\mathbf{p}}}\frac{\partial\Lambda}{\partial t}(x)\right].
\end{equation}
Two successive operations of $\dag$ must be the identity. Therefore, if eqs.~(\ref{eq:id1}) and (\ref{eq:id2}) are correct, then we must have
\begin{equation}
\left\{(2\pi)^{-3/2}\int d^{3}x e^{-ip\cdot x}
\sqrt{\frac{E_{\mathbf{p}}}{8m}}
\left[\dual{\Lambda}(x)-\frac{i}{E_{\mathbf{p}}}\frac{\partial\dual{\Lambda}}{\partial t}(x)\right]\xi(\p,\alpha)\right\}^{\dag}=a(\p,\alpha),\label{eq:ddag1}
\end{equation}
\begin{equation}
\left\{(2\pi)^{-3/2}\int d^{3}x e^{-ip\cdot x}\sqrt{\frac{E_{\mathbf{p}}}{8m}}
\dual{\zeta}(\p,\alpha)\left[-\Lambda(x)+\frac{i}{E_{\mathbf{p}}}\frac{\partial\Lambda}{\partial t}(x)\right]\right\}^{\dag}=b(\p,\alpha).\label{eq:ddag2}
\end{equation}
Although $[\dual{\Lambda}(x)\chi(\p,\alpha)]^{\dag}\neq[\dual{\chi}(\p,\alpha)\Lambda(x)]$, this does not imply that the Hermitian conjugation of the terms in the curly bracket are not equal to the particle and anti-particle annihilation operators. Expand the left-hand side using
\begin{equation}
\Lambda(x)=(2\pi)^{-3/2}\int\frac{d^{3}p}{\sqrt{2mE_{\mathbf{p}}}}\sum_{\alpha}[e^{-ip\cdot x}\xi(\p,\alpha)a(\p,\alpha)+e^{ip\cdot x}\zeta(\p,\alpha)b^{\dag}(\p,\alpha)],
\end{equation}
\begin{equation}
\dual{\Lambda}(x)=(2\pi)^{-3/2}\int\frac{d^{3}p}{\sqrt{2mE_{\mathbf{p}}}}\sum_{\alpha}[e^{ip\cdot x}\dual{\xi}(\p,\alpha)a^{\dag}(\p,\alpha)+e^{-ip\cdot x}\dual{\zeta}(\p,\alpha)b(\p,\alpha)]
\end{equation}
we find that eqs.~(\ref{eq:ddag1}) and (\ref{eq:ddag2}) are satisfied.



\label{Bibliography}
\bibliographystyle{JHEP}  
\bibliography{Bibliography}  

%
%
%
%
%
%
%
%
\end{document}